\documentclass[12pt]{iopart}
\usepackage{graphicx}

\begin{document}

\title{Stability analysis and improvement of the covariant BSSN formulation against the FLRW spacetime background}
\thispagestyle{plain} 
\author{Hidetomo Hoshino$^1$, Takuya Tsuchiya$^2$ and Gen Yoneda$^3$}

\address{$^1$ Graduate School of Fundamental Science and Engineering, Waseda University, 3-4-1 Okubo,
Shinjuku, Tokyo 169-8555, Japan}
\ead{rockfish3141@toki.waseda.jp}
\address{$^2$ Faculty of Economics, Meiji Gakuin University, 1-2-37 Shirokanedai, Minato, Tokyo 108-8636, Japan}
\address{$^3$ Department of Mathematics, School of Fundamental Science and Engineering, Waseda University, Okubo, Shinjuku, Tokyo 169-8555, Japan}
\vspace{10pt}

\begin{abstract}
In this study, we investigate the numerical stability of the covariant Baumgarte--Shapiro--Shibata--Nakamura (cBSSN) formulation against the Friedmann--Lemaître--Robertson--Walker spacetime.
To evaluate the numerical stability, we calculate the constraint amplification factor by the eigenvalue analysis of the evolution of the constraint.
We propose a modification to the time evolution equations of the cBSSN formulation for a higher numerical stability.
Furthermore, we perform numerical simulations using the modified formulation to confirm its improved stability.
\end{abstract}

%
%
%
%
%

\section{Introduction}
Advanced LIGO and Advanced VIRGO directly detected gravitational waves for the first time in 2015 \cite{Abbott2016}.
Numerical relativity significantly contributed to this achievement.
Since then, numerical relativity has continued to create catalogs of gravitational waveforms, and observations of gravitational waves are ongoing.
In addition to detecting gravitational waves, numerical relativity is expected to be useful for simulating and investigating various astrophysical phenomena.
One of the main concerns of numerical calculations is their stability. 
Numerical stability is needed to perform long-term simulations.
The Baumgarte--Shapiro--Shibata--Nakamura (BSSN) formulation \cite{Nakamura1987,Shibata1995,Baumgarte1998}, Z4 formulation \cite{Bona2002,Alic2012,Gundlach2005,Bernuzzi2010}, generalized harmonic formulation \cite{Garfinkle2002,Pretorius2005,Pretorius2005_CQG} are considered to improve the numerical stability of the Einstein equations.
As a method for evaluating numerical stability, an approach involving the eigenvalue analysis of constraint propagation equations (CPEs), which denote the time evolutions of the constraint variables, has also been proposed to help modify the time evolution equations of dynamical variables to suppress constraint violations \cite{Yoneda2001,Yoneda2002,Urakawa2022}.
This analysis is performed for each background spacetime, and it was found that the numerical stability varies depending on the specific background spacetime \cite{Urakawa2022}.
\par
Numerical relativity is also used in studies on inhomogeneous cosmology in recent years \cite{Rekier2015,Giblin2016,Bentivegna2016,Mertens2016,Daverio2017,Macpherson2017,Bolejko2017,Wang2018,Macpherson2019,Magnall2023,Munoz2023,Munoz2023_PRD}. 
It has been reported that nonlinearity is significant and cannot be ignored \cite{Giblin2016_AJ}; therefore, numerical relativity is a powerful tool to obtain approximate solutions of the Einstein equations.
In these studies, the Friedmann--Lemaître--Robertson--Walker (FLRW) spacetime is employed as a background.
\par
To carry out more and more cosmological simulations using numerical relativity, highly accurate and long-term numerical calculations are necessary.
Furthermore, since the FLRW spacetime is commonly represented using polar coordinates, a reformulation of the Einstein equations that is compatible with polar coordinates is required.
Thus, we utilize the covariant Baumgarte--Shapiro--Shibata--Nakamura (cBSSN) formulation proposed by Brown \cite{Brown2009}.
In \cite{Urakawa2022}, the numerical stability against the Minkowski, Schwarzchild, and Kerr spacetimes was studied when the cBSSN formulation was adopted.
However, we should investigate the numerical stability against the FLRW spacetime background for cosmological simulations since the numerical stability differs depending on the background spacetime.
In this paper, therefore, we carry out the eigenvalue analysis of CPEs in the cBSSN formulation assuming that the FLRW spacetime is a background.
This analysis is conducted in a spatial coordinate-free manner since we adopt the spatial covariant formulation. 
On the basis of the analysis, we modify the time evolution equations by adding a constraint variable to them with a damping parameter so that constraint violations can be suppressed. In studies on inhomogeneous cosmology, such modifications to the time evolution equations are used to improve the numerical stability \cite{Mertens2016}.
\par
In section \ref{sec:cBSSN}, we review the cBSSN formulation.
We use the same notations in \cite{Urakawa2022}.
We introduce the time evolution equations, constraint variables, and CPEs.
In section \ref{sec:CAF}, we introduce the definition of constraint amplification factor (CAF), which indicates numerical stability. We obtain CAFs against the FLRW spacetime background in section \ref{sec:CAF against FLRW}.
We find unstable modes that can cause constraint violations. We propose the way of modifying the time evolution equation of the cBSSN formulation in section \ref{sec:adjusted system against FLRW}.
We expect that constraint violations will be suppressed by this modification. We show a numerical example in section \ref{sec:numerical results}.
We compare the original cBSSN formulation and its modified version.
\par
We adopt geometric units with $G=c=1$. Greek indices run from 0
to $n$, whereas small Latin indices run from 1 to $n$.
Summation is implied over indices that appear as both upper and lower indices.
\section{Review of cBSSN formulation}
\label{sec:cBSSN}
In this section, we introduce the time evolution equations and the constraint equations of the cBSSN formulation.
We assume that the spacetime metric is written as
\begin{eqnarray}
  g_{\mu\nu}=
  \left(
      \begin{array}{cc}
       -\alpha^2+\gamma_{kl}\beta^k\beta^l & \gamma_{ik}\beta^k  \\
       \gamma_{jk}\beta^k & \gamma_{ij}  \\      
      \end{array}
  \right),
\end{eqnarray}
where $\alpha$ is the lapse function, $\beta^i$ is the shift vector, and $\gamma_{ij}$ is the spatial metric.
We define the scalar function $\phi$ as
\begin{eqnarray}
  \phi=\frac{1}{4n}\log \frac{\det (\gamma_{ij})}{\det (f_{ij})},
\end{eqnarray}
where $f_{ij}$ is any second-order positive definite tensor and $n$ denotes the spatial dimension.
We utilize $W=e^{-2\phi}$ instead of $\phi$ as a basic variable of the cBSSN formulation.
The conformal metric $\tilde{\gamma}_{ij}$ is defined as $\tilde{\gamma}_{ij}=W^2 \gamma_{ij}$.
The trace part of the extrinsic curvature $K_{ij}$ is written as $K=\gamma^{ij}K_{ij}$, and the traceless-part is $\tilde{A}_{ij}=W^2(K_{ij}-(1/n)K\gamma_{ij})$.
$K$ and $\tilde{A}_{ij}$ are basic variables of the cBSSN formulation.
We define $H^i{}_{jk}$ and $\Delta^{i}{}_{jk}$ as
\begin{eqnarray}
  H^i{}_{jk}=\frac{1}{2}h^{im}(\partial_j h_{km}+\partial_k h_{jm}-\partial_m h_{jk}),\\
  \Delta^{i}{}_{jk}=\tilde{\Gamma}^i{}_{jk}-H^i{}_{jk},
\end{eqnarray} 
where $h_{ij}$ is any second-order nondegenerate tensor and $\Delta^{i}{}_{jk}=(1/2)\tilde{\gamma}^{im}(\partial_j \tilde{\gamma}_{mk}+\partial_k \tilde{\gamma}_{mj}-\partial_m \tilde{\gamma}_{jk})$.
The vector $\bar{\Lambda}^i$ is also a new variable that satisfies $\bar{\Lambda}^i=\tilde{\gamma}^{jk}\Delta^i{}_{jk}$ at the initial time.
All basic variables $(W,\tilde{\gamma}_{ij},K,\tilde{A}_{ij},\bar{\Lambda}^i)$ for the cBSSN formulation are now prepared.
The energy-momentum tensor $T_{\mu\nu}$ is decomposed to
\begin{eqnarray}
  \rho_H &=& n^{\mu}n^{\nu}T_{\mu\nu}, \\
  J_i &=& P^{\mu}{}_i n^{\nu} T_{\mu\nu}, \\
  S_{ij} &=& P^{\mu}{}_i P^{\nu}{}_j T_{\mu\nu},
\end{eqnarray}
where $P^{\mu}{}_i=g^{\mu}{}_i +n^{\mu}n_{i},n^{\mu}=(1/\alpha,-\beta^i/\alpha)$, and $S=P^{ij}S_{ij}$.
The $n$-dimensional Ricci tensor $R_{ij}$ is decomposed as
\begin{eqnarray}
  R_{ij}
  &=& \tilde{R}_{ij} + R^W{}_{ij},\\
  \tilde{R}_{ij}
  &=&-\frac{1}{2}(\tilde{\gamma}^{mn}\mathcal{D}_m\mathcal{D}_n
    \tilde{\gamma}_{ij})
    + \tilde{\gamma}_{m(i}(\mathcal{D}_{j)}\bar{\Lambda}^m)
    \nonumber \\
    &&+ \Delta^\ell\Delta_{(ij)\ell}
    + 2\Delta^{mn}{}_{(i}\Delta_{j)mn}
    + \Delta^{mn}{}_i\Delta_{mnj},
  \\
  R^W{}_{ij}&=&
  (n-2)W^{-1}(\tilde{D}_i \tilde{D}_j W)
  +(1-n)W^{-2}(\tilde{D}_k W)(\tilde{D}^k W)\tilde{\gamma}_{ij} \nonumber
  \\
  &&+W^{-1}(\tilde{D}^k \tilde{D}_k W)\tilde{\gamma}_{ij},
\end{eqnarray}
where $\mathcal{D}_i$ is the covariant derivative operator associated with $h_{ij}$ and $\tilde{D}_i$ is the covariant derivative operator associated with $\tilde{\gamma}_{ij}$.
\par
The time evolution equations for the basic variables for the cBSSN formulation are
\begin{eqnarray}
  \partial_t W &=&
   {1\over n}\alpha WK
  -{1\over n}W(\tilde{D}_k\beta^k)
  +\beta^k (\tilde{D}_kW),
  \label{evow}
  \\
  \partial_t \tilde{\gamma}_{ij}&=&
  -2\alpha \tilde{A}_{ij}
  -{2\over n}(\tilde{D}_k\beta^k)\tilde{\gamma}_{ij}
  +2\tilde{\gamma}_{k(i}(\tilde{D}_{j)}\beta^k),
  \label{evotg}
  \\
  \partial_t K&=&
  {1\over n}\alpha K^2
  + \alpha \tilde{A}_{ij}\tilde{A}^{ij}
  -W^2(\tilde{D}^i\tilde{D}_i\alpha) \nonumber \\
  &&+(n-2)W(\tilde{D}^i\alpha)(\tilde{D}_iW)
  +\beta^i(\tilde{D}_i K)
  \nonumber \\&&
  +{2\over 1-n}\alpha \Lambda 
  +{8\pi\over n-1}\alpha S
  +{8\pi(n-2)\over n-1} \alpha \rho_H,
  \label{evok}
  \\
  \partial_t \tilde{A}_{ij}&=&
  \alpha W^2 R^{\mathrm{TF}}_{ij}
  +\alpha K\tilde{A}_{ij}
  -2\alpha\tilde{A}_{ik}\tilde{A}^k{}_{j}
  -8\pi W^2 \alpha S^\mathrm{TF}_{ij} \nonumber \\
  &&-2W\{(\tilde{D}_{(i}\alpha)(\tilde{D}_{j)}W)\}^\mathrm{TF}
  -W^2(\tilde{D}_i\tilde{D}_j\alpha)^\mathrm{TF}
  +(\tilde{D}_i\beta^k)\tilde{A}_{jk} \nonumber \\
  &&+(\tilde{D}_j\beta^k)\tilde{A}_{ik}
  + \beta^k(\tilde{D}_k\tilde{A}_{ij})
  -{2\over n}(\tilde{D}_k\beta^k) \tilde{A}_{ij},
  \label{evoa}
  \\
  \partial_t \bar{\Lambda}^i&=&
  -2(\tilde{D}_j\alpha)\tilde{A}^{ij}
    - 2n\alpha W^{-1}(\tilde{D}_jW)\tilde{A}^{ij}
    + 2\alpha \Delta^i{}_{mn}\tilde{A}^{mn} \nonumber \\
    &&- \frac{2(n-1)}{n}\alpha \tilde{\gamma}^{ij}(\tilde{D}_jK)
    - 16\pi \alpha W^2 J^i
    + \tilde{\gamma}^{mn}(\mathcal{D}_m\mathcal{D}_n\beta^i)
    \nonumber\\
    &&
    + \frac{n-2}{n}(\tilde{D}^i\tilde{D}_m\beta^m)
    + \frac{2}{n}\Delta^i (\tilde{D}_m\beta^m) \nonumber \\&&
    + (\tilde{D}_j\bar{\Lambda}^i)\beta^j
    - \bar{\Lambda}^j(\tilde{D}_j\beta^i),
  \label{evol2}
\end{eqnarray}
where $\Lambda$ is the cosmological constant and $\mathrm{TF}$ denotes the trace-free part, that is, $R^{\mathrm{TF}}_{ij}=R_{ij}-(1/n)W^{-2}\tilde{\gamma}_{ij}R=R_{ij}-(1/n)\gamma_{ij}R$.
The time evolution equations for the matter fields $\rho_H$ and $J_i$ are
\begin{eqnarray}
  \partial_t \rho_H&=&
  -\alpha W^2(\tilde{D}^k J_k)
  +(n-2)\alpha W (\tilde{D}^k W) J_k
  +\alpha K \rho_H
  \nonumber \\&&
  +\alpha W^2\tilde{A}^{mn}S_{mn}
  +{1\over n} \alpha KS
  -2(\tilde{D}^k \alpha)W^2 J_k
  \nonumber \\&&
  +\beta^k (\tilde{D}_k\rho_H),
  \\
  \partial_t J_i&=&
  -\alpha W^2(\tilde{D}^k S_{ki})
  +(n-2)\alpha W(\tilde{D}^k W)S_{k i}
  -\alpha W^{-1}(\tilde{D}_i W)S
  \nonumber \\&&
  +\alpha K J_i
  - W^2(\tilde{D}^k \alpha) S_{ki}
  -(\tilde{D}_i\alpha)\rho_H 
  \nonumber \\&&
  +\beta^k (\tilde{D}_k J_i)
  +(\tilde{D}_i \beta^k)J_k.
  \end{eqnarray}
The dynamical variables $(W,\tilde{\gamma}_{ij},K,\tilde{A}_{ij},\bar{\Lambda}^i,\rho_H,J_i)$ must satisfy the following constraint equations at each time:
\begin{eqnarray}
  {\mathcal H}
  &=&
  W^2\tilde{R}
  +(n-n^2)(\tilde{D}^k W)(\tilde{D}_k W)
  +(2n-2)W(\tilde{D}^k \tilde{D}_k W), \nonumber
  \\
  &&+{n-1\over n}K^2
  - \tilde{A}_{ij}\tilde{A}^{ij}
  -2\Lambda -16\pi \rho_H =0, \label{const_H}
  \\
  {\mathcal M}_i
  &=&
  \tilde{D}^j\tilde{A}_{ji}
  -n W^{-1}(\tilde{D}^j W)\tilde{A}_{ji}
  +{1-n\over n}(\tilde{D}_i K)
  -8\pi J_i =0,
  \label{const_M}
  \\
  {\mathcal G}^i&=&
  \bar{\Lambda}^i-\tilde{\gamma}^{jk}\Delta^i{}_{jk} =0, \label{const_G}
  \\
  {\mathcal S}&=&
  \frac{\mathrm{det}(\tilde{\gamma}_{ij})}{\mathrm{det}(f_{ij})}-1 =0, \label{const_S}
  \\
  {\mathcal A}&=&
  \tilde{A}_{ij}\tilde \gamma^{ij} =0. \label{const_A}
  \end{eqnarray}
 The CPEs are
  \begin{eqnarray}
    \partial_t \mathcal{H}&=&
  \frac{2}{n}\alpha K\mathcal{H}
  -2(\tilde{D}^j\alpha)W^2 \mathcal{M}_j
  +(2n-4)\alpha W(\tilde{D}^jW)\mathcal{M}_j
  \nonumber \\&&
  -2\alpha W^2\Delta^m{}_{m}{}^j\mathcal{M}_j
  - 2\alpha\tilde{A}^{mn}W^2 (\Delta_{mn k}\mathcal{G}^k)^\mathrm{TF}
  \nonumber \\&&
  +\frac{2}{n}\alpha K W^2\Delta^l{}_{lk}\mathcal{G}^k
  -\frac{2}{n}\alpha K W^2(\tilde{D}_k\mathcal{G}^k)
  \nonumber \\&&
  + 2\alpha \tilde{A}^{mn}W^2(\tilde{D}_m\mathcal{G}_n)^\mathrm{TF}
  +\frac{2n-2}{n}(\tilde{D}^k\tilde{D}_k\alpha)W^2\mathcal{A}
  \nonumber \\&&
  +\left(-2n+2-\frac{4}{n}\right)(\tilde{D}_k \alpha)W(\tilde{D}^kW) \mathcal{A}
  + \frac{2}{n}\alpha R \mathcal{A}
  \nonumber \\&&
  - \frac{2}{n}\alpha W^2(\tilde{D}_k\mathcal{G}^k) \mathcal{A}
  + \frac{2}{n}\alpha W^2\Delta^l{}_{lk}\mathcal{G}^k  \mathcal{A}
  \nonumber \\&&
  - \frac{16}{n}\pi  \alpha S \mathcal{A}
  +(2n-2)\alpha (\tilde{D}^kW)(\tilde{D}_kW)\mathcal{A}
  \nonumber \\&&
  -2\alpha W(\tilde{D}^k \tilde{D}_k W)\mathcal{A}
  +4(\tilde{D}^k\alpha )W^2(\tilde{D}_k\mathcal{A})
  \nonumber \\&&
  +(2-2n)\alpha W(\tilde{D}^kW)(\tilde{D}_k\mathcal{A})
  +2\alpha W^2(\tilde{D}^k\tilde{D}_k\mathcal{A}) + \beta^k (D_k\mathcal{H}), \label{CP_H}
  \\
  \partial_t \mathcal{M}_i&=&
  -\frac{1}{n}(\tilde{D}_i\alpha)\mathcal{H}
  +\frac{n-2}{2n}\alpha (\tilde{D}_i\mathcal{H})
  +\alpha K \mathcal{M}_i
  \nonumber \\&&
  +(2-n)\alpha W(\tilde{D}^m W) (\Delta_{mi k}\mathcal{G}^k)^\mathrm{TF}
  \nonumber \\&&
  +\frac{1}{n}(\tilde{D}_i\alpha)W^2\Delta^l{}_{lk}\mathcal{G}^k
  +\frac{2-n}{n}\alpha W(\tilde{D}_iW)\Delta^l{}_{lk}\mathcal{G}^k
  \nonumber \\&&
  -\frac{1}{2}\alpha W^2(\tilde{D}_i\Delta^l{}_{lk})\mathcal{G}^k
  +\alpha W^2 (\tilde{D}^m\Delta_{mi k})\mathcal{G}^k
  \nonumber \\&&
  +(\tilde{D}^m\alpha)W^2 (\Delta_{mi k}\mathcal{G}^k)^\mathrm{TF}
  +(n-2)\alpha W(\tilde{D}^m W)(\tilde{D}_m\mathcal{G}_i)^\mathrm{TF}
  \nonumber \\&&
  -\frac{1}{n}(\tilde{D}_i\alpha)W^2(\tilde{D}_k\mathcal{G}^k)
  +\frac{n-2}{n}\alpha W(\tilde{D}_iW)(\tilde{D}_k\mathcal{G}^k)
  \nonumber \\&&
  -\frac{1}{2}\alpha W^2\Delta^l{}_{lk}(\tilde{D}_i\mathcal{G}^k)
  -(\tilde{D}^m\alpha) W^2(\tilde{D}_m\mathcal{G}_i)^\mathrm{TF}
  \nonumber \\&&
  +\alpha W^2 \Delta^m{}_{i k}(\tilde{D}_m\mathcal{G}^k)
  +\frac{1}{2}\alpha W^2 (\tilde{D}_i\tilde{D}_k\mathcal{G}^k)
  -\alpha W^2 (\tilde{D}^m \tilde{D}_m\mathcal{G}_i)
  \nonumber \\&&
  -(\tilde{D}_k \alpha) \tilde{A}_{i}{}^k\mathcal{A}
  +\frac{1}{n}(\tilde{D}_i\tilde{D}_k\beta^k)\mathcal{A}
  -\alpha \tilde{A}_{i}{}^k(\tilde{D}_k \mathcal{A})
  \nonumber \\ &&
  +(\tilde{D}_i \beta^k)\mathcal{M}_k
  +\beta^k(\tilde{D}_k \mathcal{M}_i),  \label{CP_M}
  \\
  \partial_t \mathcal{G}^i&=&
  2\alpha \tilde{\gamma}^{ji} \mathcal{M}_j 
  -\alpha (\tilde{D}^i\mathcal{A})
  +\frac{2}{n}(\tilde{D}_k\beta^k)\mathcal{G}^i,  \label{CP_G}
  \\
  \partial_t \mathcal{S}&=&
  -2\alpha \mathcal{S}  \mathcal{A}
  -2\alpha  \mathcal{A}, \label{CP_S}
  \\
  \partial_t \mathcal{A}&=&
  \alpha K\mathcal{A} + \beta^k(\tilde{D}_k\mathcal{A}).  \label{CP_A}
  \end{eqnarray}
  These equations show that all constraints remain zero at all times if they are initially zero.

\section{Review of CAF and adjusted system}
\label{sec:CAF}
In this section, we review CAF, which is one of the tools for expressing the numerical stabilities of the Einstein equations.
When numerically solving time evolution equations with constraints such as the Einstein equations, it is necessary to check whether the constraints are preserved.
If the constraints are no longer satisfied, the evolving variables are far from the exact solutions.
There is a possibility that numerical errors increase exponentially.
This is because the dominant terms of the CPEs are constructed by a linear combination of the constraints. 
CAF is an indicator of the extent to which numerical errors are likely (or unlikely) to increase. CAF is calculated as follows.
\par
First, we suppose that $u^A(x^i, t)$ is a set of dynamical variables and their evolution equations are
\begin{eqnarray}
  \partial_t u^A = f^A(u^B, \ \tilde{D}_C u^B,\ \cdots).
  \label{eq:timeev}
\end{eqnarray}
Here, $A,B,\cdots$ represent the number of components of dynamical variables.
For the cBSSN formulation, $u^A=(W,\tilde{\gamma}_{ij},K,\tilde{A}_{ij},\bar{\lambda}^{i},\rho_{H},J_{i})$. 
The (first class) constraints $C^I$ satisfy
\begin{eqnarray}
  C^I(u^A,\ \tilde{D}_i u^B,\ \cdots) = 0. \label{eq:first_class_constraints}
\end{eqnarray}
Here, $I,J,\cdots$ represent the number of components of constraint variables.
Next, we suppose that the evolution equation of $C^{I}$ (called constraint propagation equations, CPEs) is written as
\begin{eqnarray}
\partial_t C^I = g^I(C^J,\ \tilde{D}_i C^J,\ \cdots),  \label{eq:CPEs_inCAF}
\end{eqnarray}
where $g^I$ is a function such as $g^I=0$ when $C^I=0$.
For the cBSSN formulation, the dominant terms of the CPEs can be expressed in the form of (\ref{eq:CPEs_inCAF}) and $C^I=(\mathcal{H},\mathcal{M}_i,\mathcal{G}^i,\mathcal{S},\mathcal{A})$. 
Finally, we apply the Fourier transformation $\tilde{D}\rightarrow \rmi\vec{k}$ to CPEs:
\begin{eqnarray}
\partial_t \hat{C}^I = \hat{g}^I(\hat{C^J}) = M^I_{\ J}\hat{C}^J,
\end{eqnarray}
where the symbol hat $\wedge$ denotes the Fourier-transformed value and $\vec{k}$ is a wave vector whose component is $(k_1,\cdots,k_n)$ and the coefficient matrix $M^I_{\ J}$ is named the constraint propagation matrix (CPM).
We call the eigenvalues of CPM the CAFs.
We modify the time evolution equation (\ref{eq:timeev}) as
\begin{eqnarray}
  \partial_t u^A = f^A(u^B, \ \tilde{D}_i u^B,\ \cdots)+ \kappa h^A(C^I, \tilde{D}_i C^J, \cdots),
\end{eqnarray}
where $h^A$ is a function whose value is 0 as long as $C^I=0$ and $\kappa$ is a parameter.
Depending on the sign of the real part of the CAF, we can predict the evolution of constraint violations: positive values imply growth, whereas negative values suggest decay. 
Even if the real part of the CAF is not negative, a smaller real part suppresses the constraint violation. The value of $\kappa$ should be decided before performing simulations so that the violations of constraints are expected to decay.

\section{CAFs against the FLRW spacetime}
\label{sec:CAF against FLRW}
In this section, we find CAFs against the FLRW spacetime.
The FLRW spacetime is represented by 
\begin{eqnarray}
  \rmd s^2=-\rmd t^2+a(t)^2 \left(\frac{\rmd r^2}{1-L r^2} + r^2 \rmd\theta^2+ r^2 \sin^2\theta \rmd\phi^2 \right),
\end{eqnarray}
where $a(t)$ is the scale factor and $L=-1,0$ or $1$, which represents the curvature of the spacetime.
We set the spatial dimension $n$ as 3 hereafter. We assume that space is flat, so we set $L=0$, and we suppose $a(t)>0,\dot{a}(t)>0$. 
Assuming this spacetime, we obtain the Fourier-transformed CPEs as
\begin{eqnarray}
  \partial_t \hat\mathcal{H}
    &=& \frac{2}{3}{K} \hat\mathcal{H}-\frac{2}{3}\alpha K W^2 (\rmi k_l\hat{\mathcal{G}}^l)-\frac{16}{3}\pi\alpha S\hat\mathcal{A}+2\alpha W^2(-|\vec{k}|^2\hat\mathcal{A}),\label{CPEH}\\
  \partial_t \hat\mathcal{M}_i
    &=& \frac{1}{6}\alpha\rmi k_i \hat\mathcal{H}+\alpha K \hat\mathcal{M}_i+\frac{1}{2}\alpha W^2(-k_ik_l\hat{\mathcal{G}}^l), \label{CPEM}\\
  \partial_t \hat\mathcal{G}^i
    &=& 2\alpha\hat\gamma^{ij}\hat\mathcal{M}_j-\rmi k^i\hat\mathcal{A}, \label{CPEG}\\
  \partial_t \hat\mathcal{S}
    &=&-2\alpha\hat\mathcal{A}, \label{CPES}\\
  \partial_t \hat\mathcal{A}
    &=&\alpha K \hat\mathcal{A},\label{CPEA}
\end{eqnarray}
where
\begin{eqnarray}
  W=\frac{1}{a(t)}, K=-\frac{3\dot{a}(t)}{a(t)}, \hat\gamma^{ij}=\mbox{diag}(1,1/r^2,1/(r^2\sin^2\theta)),
\end{eqnarray}
and $|\vec{k}|$ denotes the norm of the wave vector, that is, $|\vec{k}|^2=\tilde\gamma^{ij}k_ik_j$.
Hereafter, we use $a,\dot{a}$ instead of $W,K$.
\par
From CPEs (\ref{CPEH})--(\ref{CPEA}), we can obtain the following characteristic equation to calculate the CAFs:
\begin{eqnarray}
  \lambda (\lambda a+3\dot{a})X(\lambda)^2Y(\lambda)=0,
  \label{eq:characteristic_equation}
\end{eqnarray}
where
\begin{eqnarray}
  X(\lambda)&=&a^2\lambda ^2 +3a \dot{a} \lambda -2 |\vec{k}|^2, \\
  Y(\lambda)&=&3 a^3\lambda^3 +15 a^2 \dot{a}\lambda ^2-3a \left(|\vec{k}|^2-6 \dot{a}^2\right)\lambda-4 |\vec{k}|^2 \dot{a},
\end{eqnarray}
and $\lambda$ denotes one of the CAFs.
The solutions of $X(\lambda)=0$ are
\begin{eqnarray}
  \lambda_1&=&\frac{-3 \dot{a}-\sqrt{8 |\vec{k}|^2+9 \dot{a}^2}}{2 a}<0, \\
  \lambda_2&=&\frac{-3 \dot{a}+\sqrt{8 |\vec{k}|^2+9 \dot{a}^2}}{2 a}>0.
\end{eqnarray}
Thus, two positive CAFs appear from $X(\lambda)^2=0$.
The discriminant of $Y(\lambda)$ is
\begin{equation}
  27 a^6 \left(12 |\vec{k}|^6+75 |\vec{k}|^4 \dot{a}^2+236|\vec{k}|^2 \dot{a}^4+108 \dot{a}^6\right)>0,
\end{equation}
so $Y(\lambda)=0$ has three different real solutions. According to Vieta's formulas,
\begin{eqnarray}
  \mbox{the sum of three solutions}&=&-5\frac{\dot{a}}{a}, \\
  \mbox{the product of three solutions}&=&\frac{4 |\vec{k}|^2 \dot{a}}{3 a^3},
\end{eqnarray}
therefore, the sign of two of the three solutions is negative and that of the remaining one is positive.
Thus there are three positive eigenvalues in these CAFs.
Therefore, the cBSSN formulation has unstable modes that cause the constraint violation in the FLRW spacetime.

\section{Adjusted system for the FLRW spacetime}
\label{sec:adjusted system against FLRW}
From the Fourier-transformed CPEs (\ref{CPEH})--(\ref{CPEA}), we can write
\begin{eqnarray}
  \partial_t \hat{C}^I=M^I_{\ J}\hat{C}^J,
  \label{CPMmatrix}
\end{eqnarray}
where $\hat{C}^I=(\hat\mathcal{H},\hat\mathcal{M}_l,\hat\mathcal{G}^l,\hat\mathcal{S},\hat\mathcal{A})$ is the nine-dimensional order vector and $M^I_{\ J}$ is the ninth order square matrix. $M^I_{\ J}$ is explicitly written as
\begin{eqnarray}
  M^I_{\ J}=\left(
    \begin{array}{ccccc}
    \ast & 0 & \ast & 0 & \ast \\
    \ast & \ast & \ast & 0 & 0 \\
    0    & \ast & 0 & 0 & 0 \\
    0    & 0 & 0 & 0 & \ast \\
    0    & 0 & 0 & 0 & \ast
    \end{array}
  \right),
   \label{CPMcomponent}
\end{eqnarray}
where $*$ denotes non-zero components and $M^I_{\ J}$ is partitioned into a $5\times 5$ grid of blocks with the follwing block sizes:
\begin{itemize}
  \item first, fourth, and fifth rows: $1\times 1, 1\times 3, 1\times 3, 1\times 1, 1\times 1$
  \item second and third rows: $3\times 1, 3\times 3, 3\times 3, 3\times 1, 3\times 1$.
\end{itemize}
To investigate the eigenvalues of the CPM $M^I_{\ J}$, we examine the determinant of the matrix $M^I_{\ J}-\lambda\delta^I_{\ J}$.
If we denote the components of the matrix $M^I_{\ J}-\lambda\delta^I_{\ J}$ as $b_{IJ}$, then $\det (M^I_{\ J}-\lambda\delta^I_{\ J})$ can be written as
\begin{eqnarray}
  \det (M^I_{\ J}-\lambda \delta^I_{\ J})=\sum_K b_{IK} \tilde{b}_{IK}
\end{eqnarray}
for a fixed $I$.
Here, $\tilde{b}_{IJ}$ is the cofactor of $b_{IJ}$.
$\tilde{b}_{IJ}$ can be expressed as
\begin{eqnarray}
  \tilde{b}_{IJ}=
  \left(
    \begin{array}{ccccc}
    \ast & \ast & \ast & 0 & 0 \\
    \ast & \ast & \ast & 0 & 0 \\
    \ast & \ast & \ast & 0 & 0 \\
    0    & 0 & 0 & \ast & 0\\
    \ast    & \ast & \ast & \ast & \ast
    \end{array}
  \right). \label{CPMcofactor}
\end{eqnarray}
If we employ an adjusted system that varies the components of $M^I_{\ J}$ in (\ref{CPMcomponent}) corresponding to the non-zero components in (\ref{CPMcofactor}), the CAFs would change.
Conversely, if an adjusted system that changes the components of $M^I_{\ J}$ corresponding to the components that are 0 in (\ref{CPMcofactor}) is adopted, the CAFs do not change.
\par
For example, if we modify the time evolution equation of $W$ (\ref{evow}) to
\begin{eqnarray}
  \partial_t W=\mathrm{[original\ terms]}+\kappa {\mathcal S},
\end{eqnarray}
the Fourier-transformed CPE of ${\mathcal H}$ is changes to
\begin{eqnarray}
  \partial_t \hat\mathcal{H}
    &=& \mathrm{[original\ terms]}-4\kappa \frac{|\vec{k}|^2}{a}  {\hat\mathcal S},
\end{eqnarray}
and the other CPEs do no change.
This adjustment changes the $(1,8)$ component in $M^I_{\ J}$, but the $(1,8)$ component of $\tilde{b}_{IJ}$ in (\ref{CPMcofactor}) is 0.
Therefore, the CAFs do not change. This adjustment is of no help in terms of varying CAFs.
\par
On the other hand, we propose the time evolution equation of $\bar{\Lambda}^i$ (\ref{evol2}) be adjusted as
\begin{eqnarray}
  \partial_t \bar{\Lambda}^i=\mathrm{[original\ terms]}+\kappa {\mathcal G}^i
  \label{eq:LGadjustment}
\end{eqnarray}
to change CAFs.
If we apply this adjustment, some of the Fourier-transformed CPEs are changed as 
\begin{eqnarray}
  \partial_t \hat\mathcal{H}
    &=& \mathrm{[original\ terms]}+\kappa \alpha W^2 (k_l\hat{\mathcal{G}}^l),\label{adCPEH}\\
  \partial_t \hat\mathcal{G}^i
    &=& \mathrm{[original\ terms]}+\kappa\hat{\mathcal{G}}^i. \label{adCPEG}
\end{eqnarray}
From this, the $(1,5),(1,6),(1,7),(5,5),(5,6),(5,7),(6,5),(6,6),(6,7),(7,5),(7,6)$, and $(7,7)$ components in $M^I_{\ J}$ are changed, and these components of $\tilde{b}_{IJ}$ in (\ref{CPMcofactor}) are not 0.
We show that choosing $\kappa<0$ makes the real part of the CAFs smaller than choosing $\kappa=0$, that is, adopting the original systems. 
The characteristic equation of the CPM becomes
\begin{eqnarray}
  \bar{\lambda}  \left(3 \dot{a}+\bar{\lambda}  a\right)\bar{X}(\bar{\lambda})^2 \bar{Y}(\bar{\lambda})=0,
\end{eqnarray}
where
\begin{eqnarray}
  \bar{X}(\bar{\lambda})
    &=&X(\bar{\lambda})-\kappa  a \left(\bar{\lambda}  a+3 \dot{a}\right), \label{adjustedX}\\
    \bar{Y}(\bar{\lambda})
    &=&Y(\bar{\lambda})+\kappa  a \left(|\vec{k}|^2-3 \bar{\lambda} ^2 a^2-15 \bar{\lambda} a \dot{a}-18 \dot{a}^2\right),
    \label{eq:tildeY}
\end{eqnarray}
and $\bar{\lambda}$ denotes the CAF when the adjusted equation (\ref{eq:LGadjustment}) is applied. 
First, we consider the solutions of $\bar{X}(\bar{\lambda})=0$.
We can write the two solutions as
\begin{eqnarray}
  \bar{\lambda}=\frac{\kappa a-3\dot{a}\pm\sqrt{D}}{2a}
\end{eqnarray}
where $D=a^4\kappa^2+6a^3\dot{a}\kappa+8|\vec{k}|^2a^2+9a^2\dot{a}^2=a^4(\kappa+2\dot{a}/a)^2+8|\vec{k}|^2 a^2>0$.
We can obtain
\begin{eqnarray}
  \frac{\rmd\bar{\lambda}}{\rmd\kappa}=\frac{1}{2}\pm\frac{\kappa a+3\dot{a}}{2\sqrt{(\kappa a+3\dot{a})^2+8|\vec{k}|^2}}>0.
\end{eqnarray}
Thus, we obtain the smaller real solutions of $\bar{X}(\bar{\lambda})=0$ if we choose $\kappa<0$.
\par
Next, we consider the solutions of $\bar{Y}(\bar{\lambda})=0$.
Our goal is to show that the solutions of this equation increase monotonically with $\kappa$. 
First, we prove that all solutions of $\bar{Y}(\bar{\lambda})=0$ are real by showing $\bar{Y}(\mu_1)\bar{Y}(\mu_2)\le 0$, where $\mu_1$ and $\mu_2$ are the solutions of $\rmd\bar{Y}/\rmd\bar{\lambda}=0$.
We regard $\bar{Y}(\mu_1)\bar{Y}(\mu_2)$ as a function of $|\vec{k}|$ and $\kappa$. The maximum value of $\rmd\bar{Y}/\rmd\bar{\lambda}$ should be found either at a stationary point or at the boundary.
We solve $(\partial_{|\vec{k}|}\bar{Y}(\bar{\lambda}),\partial_{\kappa}\bar{Y}(\bar{\lambda}))=(0,0)$ to find stationary points.
The real solutions of this equation are $(|\vec{k}|,\kappa)=(0,-2\dot{a}/a),(0,-5\dot{a}/(2a))$, and $(0,-3\dot{a}/a)$.
Substituting $(|\vec{k}|,\kappa)=(0,-2\dot{a}/a),(0,-3\dot{a}/a)$ into $\bar{Y}(\mu_1)\bar{Y}(\mu_2)$, we obtain zero, and substituting $(|\vec{k}|,\kappa)=(0,-5\dot{a}/(2a))$, we obtain $-\dot{a}^6/48<0$.
When we consider the boundaries represented by the limits $|\vec{k}|\to \infty$ and $\kappa\to\pm\infty$, we find $\bar{Y}(\bar{\lambda})\to -\infty$ in all cases.
Therefore, we can say that $\bar{Y}(\mu_1)\bar{Y}(\mu_2)\le 0$ and that all solutions of $\bar{Y}(\bar{\lambda})=0$ are real.
\par
In the case where $f(\bar{\lambda})\equiv |\vec{k}|^2-3 \bar{\lambda} ^2 a^2-15 \bar{\lambda} a \dot{a}-18 \dot{a}^2$ in (\ref{eq:tildeY}) is 0, the solutions of (\ref{eq:tildeY}) should satisfy $Y(\bar{\lambda})=0$ and $f(\bar{\lambda})=0$.
However, when substituting the solutions of $f(\bar{\lambda})=0$ into $Y(\bar{\lambda})$, we obtain $(-15\dot{a}\pm\sqrt{9\dot{a}^2+12|\vec{k}|^2})/(6a)$.
These do not become zero except when $|\vec{k}|^2=0$.
Therefore, when $|\vec{k}|^2$ is not equal to 0, we do not need to consider the situation where $f(\bar{\lambda})=0$.
Furthermore, when $|\vec{k}|^2=0$, the solutions of $\bar{Y}(\bar{\lambda})=0$ are $\kappa,-2\dot{a}/a,-3\dot{a}/a$.
The first of these solutions decreases monotonically with respect to $\kappa$.
\par
In the case where $f(\bar{\lambda})\neq 0$, we can rewrite $\bar{Y}(\bar{\lambda})=0$ as
\begin{eqnarray}
  \kappa=\frac{3 a^3 \bar{\lambda} ^3 +15 a^2\dot{a}\bar{\lambda} ^2+18 a \dot{a}^2 \bar{\lambda}-3|\vec{k}|^2 a \bar{\lambda} -4 |\vec{k}|^2 \dot{a}}{a \left(3a^2\bar{\lambda} ^2+15 a \dot{a}\bar{\lambda}+18 \dot{a}^2-|\vec{k}|^2\right)}.
\end{eqnarray}
Then, we obtain
\begin{eqnarray}
  \frac{\rmd\kappa}{\rmd\bar{\lambda}}=1+\frac{2\left(3|\vec{k}|^2 (a\lambda+2\dot{a})^2+|\vec{k}|^4\right)}{\left(|\vec{k}|^2-3 (a\lambda+2\dot{a}) \left(a\lambda+3\dot{a}\right)\right)^2}>0.
\end{eqnarray}
Thus, by selecting $\kappa<0$, we obtain smaller solutions of $\bar{Y}(\bar{\lambda})=0$.
In conclusion, if we modify the evolution equation of $\bar{\Lambda}^i$ as (\ref{eq:LGadjustment}) with $\kappa<0$, we obtain smaller real CAFs than the original system.
Therefore, we expect to obtain a smaller constraint error if we set $\kappa<0$ than if we use the original system.

\section{Numerical results}
\label{sec:numerical results}
  We conduct numerical calculations under the following conditions:
  We set the background spacetime as the matter-dominated universe.
  We define the energy-momentum tensor $T_{\mu\nu}=(\rho+P)u_{\mu}u_{\nu}+Pg_{\mu\nu}$, where $u^{\mu}=(1,0,0,0)$ and $\rho$ and $P$ are the energy density and the pressure of the fluid, respectively.
  We introduce the equation of state of the matter in the universe. This equation can be written as
  \begin{eqnarray}
    P=w\rho. \label{eq:EOS}
  \end{eqnarray}
  Here, $w$ is a constant, which determines a model of the universe.
  The matter-dominated model corresponds to the case where $w=0$. From the Einstein equations or the Friedmann equation and the equation of state (\ref{eq:EOS}), we can set the line element as 
  \begin{eqnarray}
    \rmd s^2=-\rmd t^2+9t^{4/9} \left(\rmd r^2 + r^2 \rmd\theta^2+ r^2 \sin^2\theta \rmd\phi^2 \right).
  \end{eqnarray}
  and the energy-momentum tensor $T_{\mu\nu}$ as
  \begin{eqnarray}
    T_{00}&=&\frac{1}{6\pi t^2},\\
    \mbox{others}&=&0,
  \end{eqnarray}
  which we use as the initial conditions.
  These initial conditions can be specified in the variables in the cBSSN formulation as follows:
  \begin{eqnarray}
    W&=&\frac{1}{3t_0^{2/3}},\\
    \tilde{\gamma}_{ij}&=&\mbox{diag}(1,r^2,r^2\sin^2 \theta),\\
    K&=&-\frac{2}{t_0},\\
    \tilde{A}_{ij}&=&0,\\
    \bar{\lambda}^{i}&=&0,\\
    \rho_{H}&=&\frac{1}{6\pi t_0^2},\\
    J_{i}&=&0.
  \end{eqnarray}
  Here, $t_0$ denotes the initial time. We use the iterative Crank--Nicolson method with two iterations \cite{Teukolsky2000} and utilize the exact solution for the boundary in the $r$- and $\theta$-directions and the periodic boundary condition in the $\phi$-direction.
  We set $5\le r \le 15, \pi/6 \le \theta \le 5\pi/6, 0 \le \phi \le 2\pi$, and $\Delta t=0.0125$.
  The numbers of grid points are 120 in the $r$-direction, 40 in the $\theta$-direction, and 80 in the $\phi$-direction, and the initial time $t_0=1$ and the termination time is 100. We define the constraint error ${C}$ as
  \begin{eqnarray}
    {C}=\int \sqrt{\mathcal{H}^2+\tilde{\gamma}^{ij}\mathcal{M}_i \mathcal{M}_j+\tilde{\gamma}_{ij}\mathcal{G}^i\mathcal{G}^j+\mathcal{A}^2+\mathcal{S}^2} \, \rmd V,
  \end{eqnarray}
  where $\rmd V=r^2\sin\theta dr d\theta d\phi$.
  \begin{figure}[htbp]
    \centering
      \includegraphics[keepaspectratio, scale=0.5]{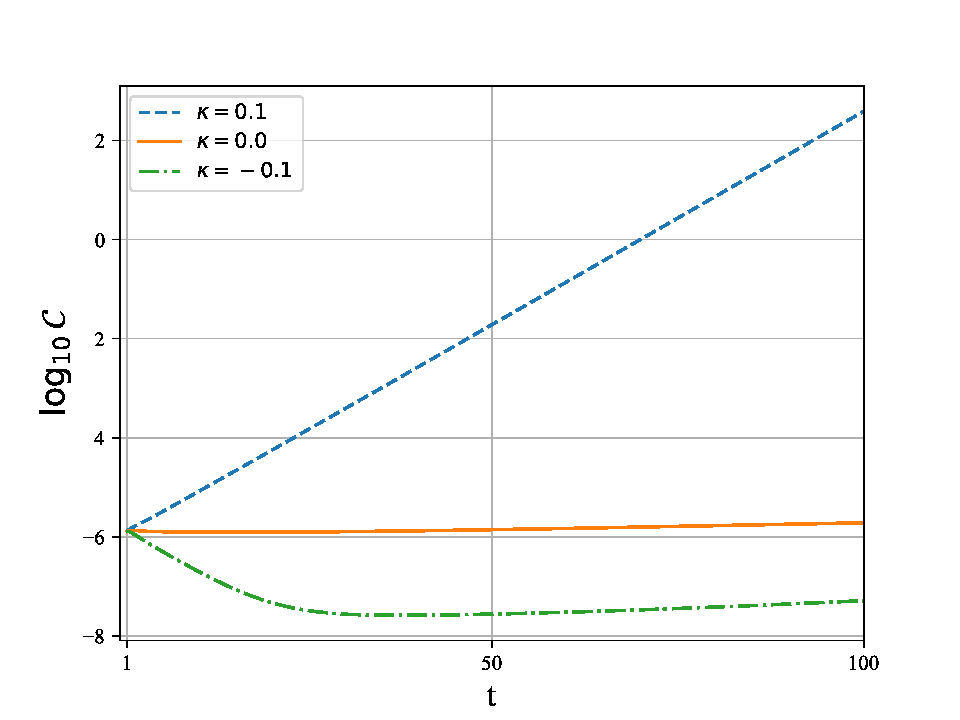}
      \caption{The horizontal axis represents time and the vertical axis represents the value of $\log_{10}{C}$. The value with $\kappa=-0.1$ is the smallest of the three.}
      \label{fig:figure_1}
  \end{figure} 
  We conduct numerical calculations for $\kappa=0.1,0,-0.1$, the results of which are shown in figure \ref{fig:figure_1}.
  The $\kappa<0$ case has the smallest constraint error and the $\kappa>0$ case has the largest constraint error.
  These results are consistent with the expected outcomes from section \ref{sec:adjusted system against FLRW}. 
  Therefore, it is found that the adjustment (\ref{eq:LGadjustment}) with a negative damping parameter $\kappa$ helps suppress the constraint violations and improve the numerical stability.
  \par
  The numerical calculation results corresponding to the vacuum-dominated universe and radiation-dominated universes are shown in  \ref{app:vacuum} and \ref{app:radiation}, respectively.
  In any calculations, we obtain the smallest constraint error when $\kappa<0$ and the largest constraint error when $\kappa>0$.
  This is consistent with the fact that, regardless of the cosmological model $a$, CAF becomes smaller when $\kappa<0$.

  \section{Summary}
  We investigated the numerical stability of the covariant Baumgarte--Shapiro--Shibata--Nakamura (cBSSN) formulation against the Friedmann--Lemaître--Robertson--Walker (FLRW) spacetime.
  To estimate the numerical stability, we performed the eigenvalue analysis of the constraint propagation equations (CPEs), which  describe the time evolution of constraint variables. 
  The results of this eigenvalue analysis yield the constraint amplification factors (CAFs), which are the eigenvalues of the coefficient matrix of CPEs and serve as indicators of numerical stability. 
  If there are positive real parts of the CAFs, the systems have unstable modes that can cause constraint violation. 
  We found that there are three positive CAFs in the cBSSN formulation against the FLRW spacetime.
  Therefore, the cBSSN formulation has unstable modes that cause the constraint violation in the homogeneous and isotropic spacetime.
  \par
  We proposed a way of adjusting the time evolution equation of the cBSSN formulation to improve the numerical stability. 
  We showed that using the adjusted time evolution equation, the real parts of the CAFs are decreased. 
  As a result, the constraint violation is expected to be suppressed. 
  We performed numerical experiments using the adjusted cBSSN formulation for the FLRW spacetime in three cases: the matter-dominated, vacuum-dominated, and radiation-dominated universes. 
  These results are consistent with the analytical prediction using CAFs.
  \par
  Our findings contribute to the ongoing efforts in using numerical relativity to improve the stability of simulations, particularly in the context of inhomogeneous cosmology. 
  In studies of inhomogeneous cosmology using numerical relativity, the background spacetime is often described by a perturbed FLRW spacetime.
  However, numerical simulations against the FLRW spacetime background can be challenging owing to the potential for numerical instabilities. 
  Our study, which focuses on the numerical stability of the cBSSN formulation against the FLRW spacetime background, provides a foundation for understanding and mitigating these instabilities.
  If the cBSSN formulation is adopted, our proposed adjustment will contribute to improving numerical stability when the spacetime is not far from the FLRW spacetime. 
  Performing numerical calculations and simulating astrophysical phenomena by introducing specific perturbations to the FLRW spacetime using our proposed adjustment is future work.

  \ack
  TT and GY were partially supported by JSPS KAKENHI Grant No.24K06856.
  TT was partially supported by JSPS KAKENHI Grant Numbers JP21K03354 and JP24K06855.
  GY was partially supported by a Waseda University Grant for Special Research Projects 2024C-107, 2023C-091, and 2023Q-008.

\appendix
\section{Numerical results: vacuum-dominated universe}
\label{app:vacuum}
The vacuum-dominated universe model corresponds to the case with $w=-1$ in the equation of state (\ref{eq:EOS}) and we set the cosmological constant as $\Lambda=3$.
We set the line element as
  \begin{eqnarray}
    \rmd s^2=-\rmd t^2+\rme^{2t} \left(\rmd r^2 + r^2 \rmd\theta^2+ r^2 \sin^2\theta \rmd\phi^2 \right)
  \end{eqnarray}
and the energy-momentum tensor as $T_{\mu\nu}=0$ as the initial conditions. 
These initial conditions can be specified in variables in the cBSSN formulation as
  \begin{eqnarray}
    W&=&e^{-t_0},\\
    \tilde{\gamma}_{ij}&=&\mbox{diag}(1,r^2,r^2\sin^2 \theta),\\
    K&=&-\sqrt{3},\\
    \tilde{A}_{ij}&=&0,\\
    \bar{\Lambda}^{i}&=&0,\\
    \rho_{H}&=&0,\\
    J_{i}&=&0.
  \end{eqnarray}
  Other settings are the same as those in section \ref{sec:numerical results}.
  \begin{figure}[htbp]
    \centering
      \includegraphics[keepaspectratio, scale=0.5]{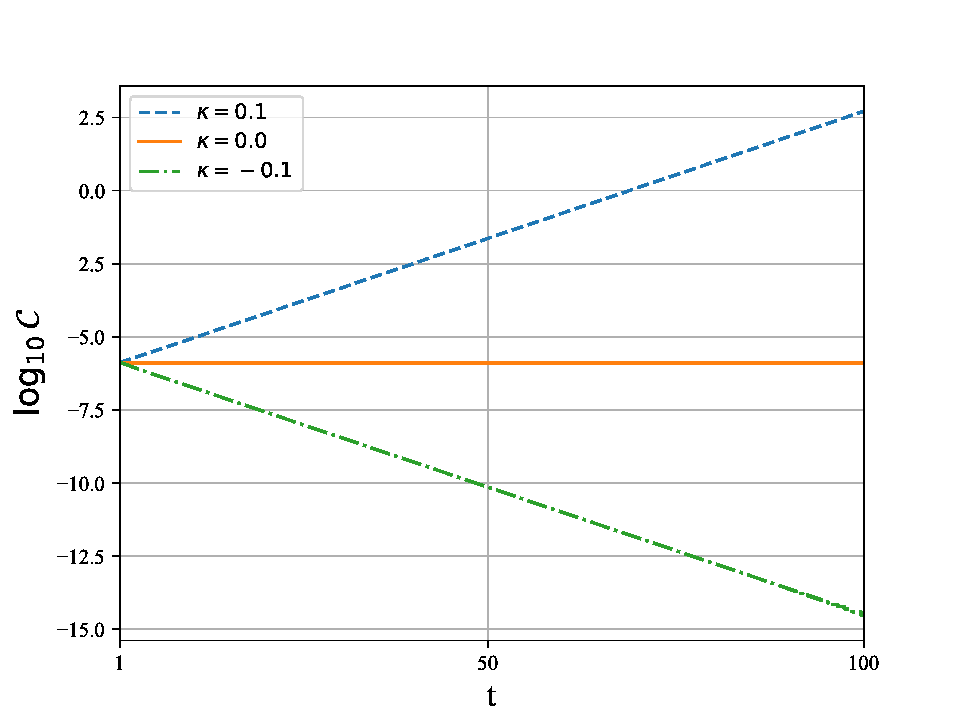}
      \caption{The horizontal axis represents time and the vertical axis represents the constraint error. The error with $\kappa=-0.1$ is the smallest of the three.}
      \label{fig:figure_a1}
  \end{figure}
  We conduct numerical calculations for $\kappa=0.1,0,-0.1$, the results of which are shown in Figure \ref{fig:figure_a1}.
  The $\kappa<0$ case has the smallest constraint error and the $\kappa>0$ case has the largest constraint error.

\section{Numerical results: radiation-dominated universe}
\label{app:radiation}
The radiation-dominated universe model corresponds to the case with $w=1/3$ in the equation of state (\ref{eq:EOS}).
We set the line element as 
\begin{eqnarray}
  \rmd s^2=-\rmd t^2+t \left(\rmd r^2 + r^2 \rmd\theta^2+ r^2 \sin^2\theta \rmd\phi^2 \right)
\end{eqnarray}
and the energy-momentum tensor $T_{\mu\nu}$ as
\begin{eqnarray}
  T_{00}&=&\frac{3}{32\pi t^2},\\
  T_{i0}&=&0, \\
  T_{ij}&=&\frac{1}{32\pi t}\mbox{diag}(1,r^2,r^2\sin^2 \theta),
\end{eqnarray}
which we use as the initial conditions. 
These initial conditions can be specified in variables in the cBSSN formulation as
\begin{eqnarray}
  W&=&\frac{1}{\sqrt{t_0}},\\
  \tilde{\gamma}_{ij}&=&\mbox{diag}(1,r^2,r^2\sin^2 \theta),\\
  K&=&-\frac{3}{2t_0},\\
  \tilde{A}_{ij}&=&0,\\
  \bar{\lambda}^{i}&=&0,\\
  \rho_{H}&=&\frac{3}{32\pi t^2_0},\\
  J_{i}&=&0.
\end{eqnarray}
Other settings are the same as those in section \ref{sec:numerical results}.
\begin{figure}[htbp]
  \centering
    \includegraphics[keepaspectratio, scale=0.5]{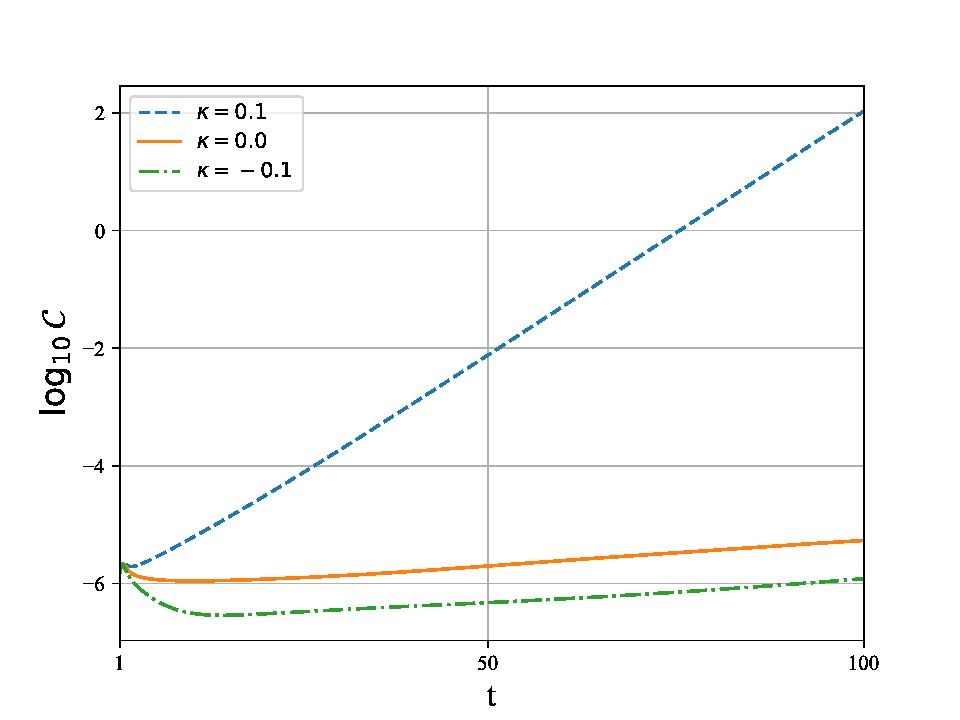}
    \caption{The horizontal axis represents time and the vertical axis represents the constraint error. The error with $\kappa=-0.1$ is the smallest of the three.}
    \label{fig:figure_b1}
\end{figure}
We conduct numerical calculations for $\kappa=0.1,0,-0.1$, the results of which are shown in figure \ref{fig:figure_b1}.
The $\kappa<0$ case has the smallest constraint error and the $\kappa>0$ case has the largest constraint error.

\section*{References}
\bibliographystyle{iopart-num}
\bibliography{myref}
\clearpage

\end{document}